\documentclass[twocolumn,amsmath,amssymb,floatfix,prb,showpacs,superscriptaddress]{revtex4}
\usepackage{amsmath,amssymb,natbib,bm,graphicx,url,epsfig}
\usepackage[ansinew]{inputenc}

\newcommand{\be}{\begin{equation}}
\newcommand{\ee}{\end{equation}}
\newcommand{\bes}{\begin{equation}\begin{split}}
\newcommand{\ees}{\end{split}\end{equation}}

\newcommand{\vc}[1]{\mathbf{#1}}

\newcommand{\abs}[1]{\left|#1\right|}

\newcommand{\ket}[1]{\left|\, #1 \, \right\rangle}

\newcommand{\boket}[3]{\left\langle\, #1 \,\left|\, #2 \,\right|\, #3 \,\right\rangle}

\DeclareMathOperator{\sgn}{sgn}

\begin{document}
\title{Current-induced nonequilibrium vibrations in single-molecule devices}
\author{Jens Koch}
\author{Matthias Semmelhack}
\author{Felix \surname{von Oppen}}
\affiliation{Institut f\"ur Theoretische Physik, Freie Universit\"at Berlin, Arnimallee 14, 14195 Berlin, Germany}
\author{Abraham Nitzan}
\affiliation{School of Chemistry, The Sackler Faculty of Science, Tel Aviv University, Tel Aviv 69978, Israel}
\date{January 31, 2006}
\begin{abstract}

Finite-bias electron transport through single molecules generally induces nonequilibrium molecular vibrations (phonons). By a mapping to a Fokker-Planck equation, we obtain analytical scaling forms for the nonequilibrium phonon distribution in the limit of weak electron-phonon coupling $\lambda$  within a minimal model. Remarkably, the width of the 
phonon distribution diverges as $\sim\lambda^{-\alpha}$ when the coupling decreases,  
with  voltage-dependent, non-integer exponents $\alpha$. This implies a breakdown of perturbation theory in the electron-phonon coupling for fully developed nonequilibrium. We also discuss possible experimental implications of 
this result such as current-induced dissociation of  molecules.

\end{abstract}
\maketitle
\section{Introduction}
The vision of molecular electronics\cite{aviram}
in part depends on the realization of devices such as molecular transistors, switches, or diodes. One strategy towards this goal involves the coupling of electronic 
and vibrational (phononic) degrees of freedom of molecules. Experiments with single-molecule devices have demonstrated effects of electron-phonon coupling in current-voltage characteristics ($IV$s),\cite{park,ruitenbeek,yu} and a number of theoretical studies have investigated such features in $IV$s,\cite{mccarthy,aleiner,braig,nitzan2,zwolak,koch2} shot noise,\cite{aleiner,koch2} and the thermopower\cite{koch} as well as applications such as diodes\cite{kaat} and switches.\cite{domcke}

A question of principal importance for single-molecule devices are the consequences of nonequilibrium effects 
at finite bias. 
Strong nonequilibrium molecular vibrations can be beneficial in molecular devices, e.g., by enhancing 
switching rates between molecular conformations. In other instances, they may hinder the operation of
devices, in the extreme case by inducing dissociation of the molecule. Recent theoretical work shows
that even within simple models, vibrational nonequilibrium has important effects on $IV$s and shot noise,\cite{varma,aleiner} may induce a shuttling instability,\cite{fedorets,novotny} or lead to current flow 
characterized by a self-similar hierarchy of avalanches of large numbers of transferred electrons.\cite{koch2}

Recent numerical results by Mitra et al.\cite{aleiner} suggest that intriguingly, vibrational nonequilibrium becomes stronger as the electron-phonon coupling $\lambda$ {\it decreases}. Characterizing the vibrational nonequilibrium by the probability distribution of phonon excitations, these authors observe that the width 
of this distribution grows with decreasing coupling $\lambda$. These numerical results are obtained within a minimal
model describing transport through one molecular orbital, which is coupled to a single vibrational mode. 

In this paper, we first clarify the underlying mechanism for this nonequilibrium effect by developing an analytical theory. Our approach relies on a mapping to a Fokker-Planck equation, which 
becomes exact in the limit of weak electron-phonon coupling. This mapping predicts that the width at half maximum (WHM) of the phonon distribution diverges as $\lambda \to 0$. 
Remarkably, the WHM is shown to scale as $\lambda^{-\alpha}$ with bias-dependent, non-integer exponents 
$\alpha>0$. We confirm our analytical results by numerical simulations. 

``Real" single-molecule devices will typically involve features such as several vibrational modes, anharmonic vibrational potentials, and direct vibrational relaxation (e.g.\ due to radiation or interaction with the substrate), which are not fully captured by the minimal model. We therefore discuss how various such extensions of the minimal model, which may be important for an accurate description of experimental systems, modify our analytical findings. Specifically, we include anharmonic vibrations within a Morse-potential model which allows us to discuss current-induced dissociation of the molecule. In this context, we show that the current-induced dissociation rate is governed by an interplay of the above-mentioned  divergence of the width of the phonon distribution and a slowing down of the diffusion in phonon space as $\lambda$ decreases.

The outline of the paper is as follows: In Sec.~\ref{sec1} we discuss the nonequilibrium effects of weak electron-phonon coupling within the minimal model. The model is specified in \ref{sec11}, and the resulting properties of the phonon distribution are derived in \ref{sec12}. Nonequilibrium properties of ``real" molecules are discussed in Sec.~\ref{sec2} by going beyond the minimal model. In particular, we address the effects of vibrational relaxation and the presence of several phonon modes as well as the situation of anharmonic potentials in \ref{sec21}, \ref{sec21a} and \ref{sec22}, respectively. Our conclusions are summarized in Sec.~\ref{sec3}.

\section{Weak electron-phonon coupling within the minimal model\label{sec1}}
\subsection{Reduced model\label{sec11}}

We investigate the nonequilibrium vibrational properties of a molecule coupled to metallic source and drain electrodes under finite bias within the following minimal model.\cite{glazman2,schoeller3,aleiner,koch} Electronic transport is taken to result from sequential tunneling through one spin-degenerate molecular orbital with energy $\varepsilon$, which is measured relative to the zero-bias Fermi energies of the leads and which can be tuned by a gate voltage. As a minimal model, we consider a single vibrational mode with frequency $\omega$. [Typical vibrational energies in molecules are of the order of 0.1 eV.]  The system's Hamiltonian reads $H= H_\text{mol} + H_\text{leads} + H_\text{mix}$, where 
\begin{align}
H_\text{mol}= &\varepsilon n_d + \frac{U}{2} n_d(n_d-1)
 \\\nonumber
&+ \lambda \hbar\omega (b^\dag + b)n_d + \hbar\omega(b^\dag b+1/2)  \label{Hmol}
\end{align}
describes the molecular degrees of freedom,  $H_\text{leads}$ a free electron gas in the leads $a=L,R$ (with creation operators $c^\dag_{a\vc{p}\sigma}$), and 
\be H_\text{mix}= \sum_{a=L,R}\sum_{\vc{p},\, \sigma} \left( t_a c^\dag_{a\vc{p}\sigma} d_\sigma + \text{h.c.}\right)
\ee
 the tunneling between leads and molecule. 

We focus on the regime of strong Coulomb blockade, appropriate when voltage and temperature are small compared to the charging energy  $U$. The operator $d_\sigma$ ($d_\sigma^\dag$) annihilates (creates) an electron with spin projection $\sigma$ on the molecule, $n_d=\sum_\sigma d_\sigma^\dag d_\sigma$ denotes the corresponding occupation-number operator. Vibrational excitations are annihilated (created) by $b$ ($b^\dag$). The electron-phonon coupling term can be eliminated by a canonical transformation,\cite{aleiner,glazman2}
leading to a renormalization of the parameters $\varepsilon$ and $U$, and of the lead-molecule coupling $t_a\rightarrow t_a \exp[-\lambda(b^\dagger-b)]$. From now on, we refer to the renormalized parameters as $\varepsilon$ and $U$. 

The coupling between molecule and leads is parameterized by the tunneling matrix elements $t_L$ and $t_R$, and it is assumed to be weak in the sense that the energy broadening $\Gamma$ of molecular levels is small, i.e.~$\Gamma\ll k_BT,\hbar\omega$, so that a perturbative treatment for $H_\text{mix}$ in the framework of rate equations, as introduced in the context of Coulomb blockade phenomena,\cite{beenakker} is appropriate. We focus on temperatures $k_B T\ll \hbar\omega$ (corresponding to typical low-temperature experiments, see e.g.~Ref.~\onlinecite{ruitenbeek}).
For simplicity, we assume a symmetric device with $t_L=t_R\equiv t_0$ and identical voltage drops of $V/2$ across each junction.\footnote{We emphasize that our assumption of a symmetric device is not crucial for our essential results. Specifically, the scaling behavior, Eq.~\eqref{scaling}, is not sensitive to asymmetries of the molecule-lead coupling.}

Then, the occupation probability $P^n_q$ for the molecular state $\ket{n,q}$ with $n$ electrons and $q$ phonons is determined by the rate equations 
\be\label{rateeq} 
\frac{dP^n_q}{dt}= \sum_{n',q'}  \left[ P^{n'}_{q'} W^{n'\rightarrow n}_{q'\rightarrow q} - P^{n}_{q} W^{n\rightarrow n'}_{q\rightarrow q'} \right].
\ee  
(The discussion of direct phonon relaxation is deferred until later in this paper.)
The transition rates $W^{n\rightarrow n'}_{q\rightarrow q'}$ obtained by Fermi's golden rule are proportional to the square of the Franck-Condon (FC) matrix element
\be\label{HO_matrixelement}
M_{q\rightarrow q'}=\int_{-\infty}^\infty dx\, \phi_{q'}(x) \phi_q(x-\sqrt{2}\lambda\ell_\text{osc}),
\ee
i.e.~the overlap of two harmonic oscillator wavefunctions $\phi_q(x)$, shifted relative to each other by a distance $\sqrt{2}\lambda\ell_\text{osc}$, with electron-phonon coupling strength $\lambda$, and vibrational oscillator length $\ell_\text{osc}=(\hbar/M\omega)^{1/2}$. Based on spectral data for diatomic molecules,\cite{herzberg} one can find electron-phonon coupling strengths ranging between $\lambda\simeq0.01$ (BeO) and $\lambda\simeq5.4$ (Kr$_2$). The FC matrix elements can be expressed as 
\be\label{lag}
M_{q\rightarrow q'}= \left( q!/Q! \right)^{1/2}\\
\lambda^{Q-q}e^{-\lambda^2/2} \, L_{q}^{Q-q}\left(\lambda^2 \right)\sgn(q-Q)^{q-Q},
\ee
see e.g.\ Ref.\ \onlinecite{koch}, where $L^n_m(x)$ denotes the generalized Laguerre polynomial.

\subsection{Phonon distributions for weak electron-phonon coupling\label{sec12}}
For $\lambda\ll1$, Eq.~\eqref{lag} 
leads to
\be\label{asymptotic}
\abs{M_{q_1\rightarrow q_2}}^2 \simeq  \frac{Q!}{q!}\frac{\lambda^{2\Delta q}}{(\Delta q!)^2},
\ee
valid for $q\lambda^2,\,\Delta q\lambda^2\ll1$, where $Q=\max\{q_1,q_2\}$, $q=\min\{q_1,q_2\}$, and $\Delta q=Q-q$.
Therefore, the FC matrix elements and the transition rates  $W^{n\rightarrow n\pm 1}_{q\rightarrow q'}$ decay rapidly with increasing  $\Delta q$. Consequently, the vibrational state of the molecule is predominantly changed by processes for which  $q\rightarrow q'=q\pm1$, and $\Delta q=1$. Neglecting all other processes, Eq.~\eqref{rateeq} describes a random walk in the space of phonon states $q$ with $q$-dependent nearest-neighbor hopping rates. In this approximation, the random walker would eventually escape to infinity, as  the rates for $q\rightleftharpoons q+1$ are equal and grow with $q$. This implies that there is \emph{no steady-state} phonon distribution $P_q=\sum_n P^n_q$  within this random-walk model.

To derive the actual steady-state phonon distribution, it is therefore imperative to go beyond the random-walk 
model by including higher-order processes with $\Delta q>1$. These may favor vibrational de-excitation processes
since the applied voltage sets an upper limit to the increase (but not to the decrease!) in the vibrational 
excitation $q$ by a tunneling event.
For example, for $\epsilon=0$ the full voltage drop $eV/2$ per sequential-tunneling event can be converted 
into vibrational energy. Thus, $\Delta q_a=\lfloor eV/2\hbar\omega \rfloor+1$ is  the leading-order asymmetric process 
{\it for which only de-excitation processes are permitted}. (Here, $\lfloor r \rfloor$ denotes the largest integer smaller or equal to $r$.) 

\begin{figure}
  \centering
		\includegraphics[width=0.7\columnwidth]{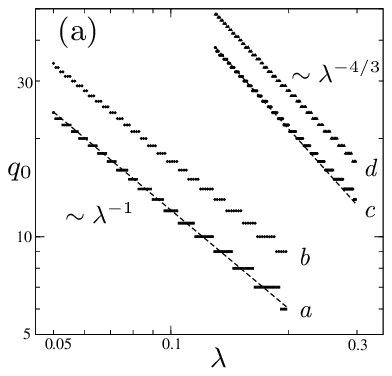}\\
		\includegraphics[width=0.7\columnwidth]{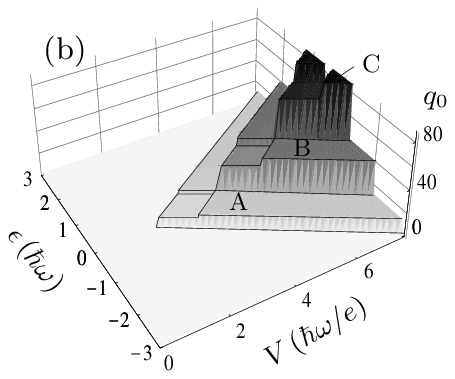}
	\caption{(a) Scaling behavior of the phonon distribution width as a function of electron-phonon coupling for representative values of $(eV/\hbar\omega,\epsilon/\hbar\omega)$ equal to \emph{a}: $(3,0)$; \emph{b}: $(5,1)$; \emph{c}: $(5,0)$; \emph{d}: $(7,1)$. (b) Width of the phonon distribution as a function of bias and gate voltage at $\lambda=0.15$ and $k_BT=0.005\hbar\omega$. The scaling of the width differs for the plateaus according to A: $q_0\sim\lambda^{-1}$, B: $q_0\sim\lambda^{-4/3}$, and C: $q_0\sim\lambda^{-3/2}$.
 \label{fig1}}	
\end{figure}

We can now derive the scaling of the steady-state phonon distribution $P_q$
with electron-phonon coupling $\lambda$ by
balancing the diffusion process due to tunneling events with $\Delta q =1$ [with diffusion constant 
$\sim q\lambda^2$, see Eq.\ (\ref{asymptotic})] and the leading asymmetric drift process [with rate  
$(q\lambda^2)^{\Delta q_a}$, see Eq.\ (\ref{asymptotic})]. This leads to the balance equation 
\be 
q\lambda^2{P_q}''\sim (q\lambda^2)^{\Delta q_a}
{P_q}'
\ee 
which implies a scaling law for the width $q_0$ of $P_q$, namely 
\be
q_0 \sim \lambda^{-\alpha},\qquad \alpha= {2(\Delta q_a-1)}/{\Delta q_a}.\label{scaling}
\ee
This power-law scaling is nicely confirmed by numerical results for $P_q$ as shown in Fig.~\ref{fig1}(a). Remarkably, for $k_B T\ll \hbar\omega$ the discrete dependence of the leading asymmetric process on bias and gate voltage implies finite regions in the $(V,\epsilon)$-plane characterized by certain {\it non-integer} exponents $\alpha$. This ``phase diagram" is shown in Fig.\ \ref{fig1}(b) where the wedge-shaped regions A, B, and C correspond to $\alpha = 1, 4/3$, 
and $3/2$, respectively.\footnote{Additional smaller steps can be traced back to changes in the nature of the 
asymmetry within one scaling phase.}

\begin{figure}
  \centering
		\includegraphics[width=0.7\columnwidth]{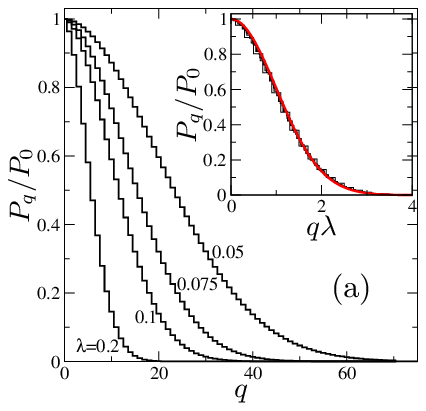}\\[0.2cm]
		\includegraphics[width=0.7\columnwidth]{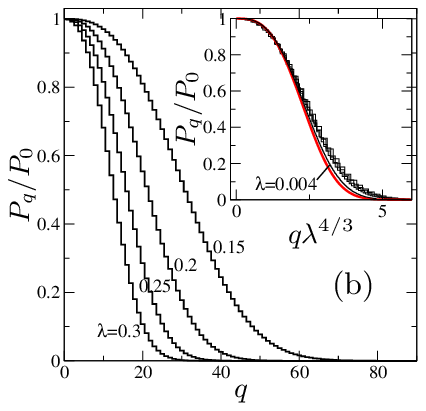}\\[0.2cm]
	\caption{(Color online) Phonon distributions $P_q$  at bias voltages (a) $eV=3\hbar\omega$ and (b) $eV=5\hbar\omega$, plotted for several coupling strengths $\lambda$ at $\epsilon=0$, $k_BT=0.05\hbar\omega$. 
The insets show that these distributions (approximately) collapse to universal curves given by the red (dark gray) smooth curves, which are the solutions of the the Fokker-Planck equation 
\eqref{FPE} specific to each voltage range $n\hbar\omega<eV/2<(n+1)\hbar\omega$.
 \label{fig2}}	
\end{figure}

For the diamond-shaped regions along the line $\epsilon=0$ in Fig.~\ref{fig1}(b), we can go beyond the derivation
of this scaling behavior and obtain analytical results for the entire phonon distribution $P_q$
by a mapping to a Fokker-Planck equation. The derivation exploits the crucial observation that 
for $U\to \infty$ and $k_BT\ll \hbar\omega$, 
the transition rates $W^{n\rightarrow n'}_{q\rightarrow q'} = s^{n\rightarrow n'} w_{q\rightarrow q'}$  
factorize into a 
spin factor $s^{n\rightarrow n'}=(1+\delta_{n',1})\delta_{\abs{n-n'},1}$ and a phonon factor
\be w_{q\rightarrow q'}=\tau_0^{-1}\abs{M_{q\rightarrow q'}}^2[ \theta(q+\Delta q_a-1-q')
+\theta(q-q'-\Delta q_a)],\ee
 where $\tau_0^{-1}= \Gamma/\hbar =
2\pi\rho\abs{t_0}^2/\hbar$.
In the stationary case, 
this implies the factorization $P^n_q=P^nP_q$, which allows us to derive the {\it purely 
phononic} rate equation 
\be
0=dP_q/dt=\sum_{q'}[P_{q'}w_{q'\rightarrow q}-P_qw_{q\rightarrow q'}],
\ee 
Since the phonon distribution becomes wide, we can take $q$ to be
continuous, expand $P_{q'}$ around $q'=q$ up to second order, and keep only the leading-order contributions to 
diffusion and drift. In this way we obtain the  Fokker-Planck equation 
\begin{align}
0=\frac{\partial P}{\partial t}=
\frac{1}{2}\frac{\partial^2}{\partial q^2}\left[  D(q) P(q) \right]
- \frac{\partial}{\partial q}\left[A(q) P(q) \right],  
\label{FPE}
\end{align}
with diffusion coefficient $D(q)=2q\lambda^2/\tau_0$, drift coefficient $A(q)=[\lambda^2-c(q\lambda^2)^{\Delta q_a}]/\tau_0$, and $c=2\Delta q_a(\Delta q_a! )^{-2}$.
Remarkably, the stationary Fokker-Planck equation  (\ref{FPE}) can be solved analytically for any $\Delta q_a$ 
by the scaling ansatz $P_q=a\lambda^\alpha f(\lambda^\alpha q)$ with  a normalization constant $a$.
The universal function $f$ is uniquely determined by Eq.~\eqref{FPE} together with the boundary 
conditions $f(0)=1$, $f'(0)=0$, and we find 
\be\label{universalf}
f(x)=\exp \left[ -x^{\Delta q_a}/b\right]
\ee
with $b=\frac{1}{2}(\Delta q_a!)^2$. Note that in particular, this analytical result confirms the power-law
scaling (\ref{scaling}) of the width of the phonon distribution. The power-law scaling (\ref{scaling}) together with the analytical phonon distributions (\ref{universalf})
constitute the central results of this paper. The phonon distributions are nicely confirmed by 
numerical solutions of the full rate equations as shown in Fig.~\ref{fig2}.\footnote{The small deviations 
observed in the inset of Fig.~\ref{fig2}(b) reflect that the effective perturbation parameter $q_0\lambda^2\sim\lambda^{2/(\Delta q_a)}$ grows with $\Delta q_a$ even at fixed $\lambda$.}

In fully developed nonequilibrium, the width of the phonon distribution \emph{diverges} 
with \emph{decreasing} electron-phonon coupling $\lambda$, and the resulting phonon distributions are 
{\it non-analytic} in $\lambda$. An important theoretical implication of this result is that in fully developed nonequilibrium, perturbation theory in the electron-phonon coupling parameter $\lambda$ is inadequate. 
Indeed, we find below that the radius of convergence of such an expansion in $\lambda$ would involve the direct vibrational relaxation rate.

\section{Implications for ``real" molecules\label{sec2}}

Transport through ``real" molecules will typically involve physics that goes beyond the minimal model. In particular, we will discuss (i) direct vibrational relaxation, (ii) the presence of more than one vibrational mode, and (iii) anharmonic vibrational potentials.  We show that while the exact scaling results for phonon distributions are specific to the minimal model, nonequilibrium effects at weak electron-phonon coupling persist and can be understood within the phonon diffusion model as long as the vibrational relaxation rate remains small compared to $I/e$.

\subsection{Direct vibrational relaxation\label{sec21}}
Direct phonon relaxation can be included within the relaxation-time approximation by adding $-\frac{1}{\tau}[{\textstyle P^n_q- P^\text{eq}_q \sum_{q'} P^n_{q'}}]$ to the r.h.s.\ of the rate equations \eqref{rateeq}. Here, $P^\text{eq}_q$ denotes the equilibrium phonon distribution, which can be approximated by $P^\text{eq}_q=\delta_{q,0}$ for $k_BT\ll\hbar\omega$. 

\begin{figure}
  \centering
		\includegraphics[width=0.6\columnwidth]{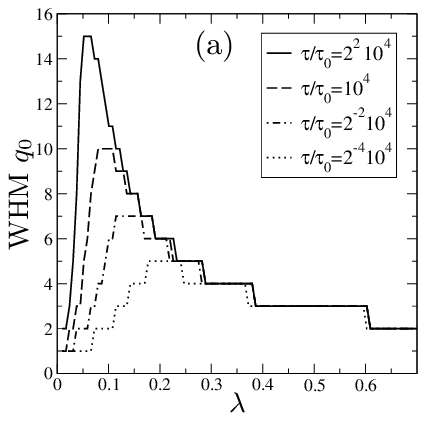}\\[0.5cm]
		\includegraphics[width=0.6\columnwidth]{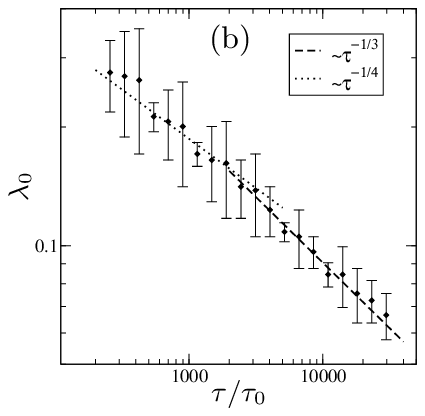}
	\caption{(a) Width of the phonon distribution as a function of electron-phonon coupling strength for several vibrational relaxation times $\tau$ and $eV=3\hbar\omega$, $k_BT=0.05\hbar\omega$, $\epsilon=0$. For $\lambda>0.25$ all curves show the approximate $q_0\sim\lambda^{-1}$ scaling. Below a relaxation-rate dependent crossover-point $\lambda_0$ the WHM is strongly suppressed due to direct relaxation. (b) Crossover-point $\lambda_0$ vs. relaxation time $\tau$.
 \label{fig3}}	
\end{figure}

To understand the effect of direct vibrational relaxation on the phonon distribution $P_q$, it is important to 
note that the diffusion and drift processes in phonon space slow down as the electron-phonon
coupling $\lambda$ decreases. As $\lambda$ decreases, we therefore expect that there exists a crossover coupling 
$\lambda_0$: For $\lambda\gg\lambda_0$, the vibrational diffusion is limited by the drift in phonon space induced 
by the asymmetry between vibrational excitation and de-excitation, as discussed above. By contrast, for $\lambda\ll \lambda_0$, the dominant limiting process is direct vibrational relaxation, leading to a decrease of the width 
of the phonon distribution. This expectation is confirmed by
numerical results as seen in Fig.\ \ref{fig3}(a) which shows the width of the phonon distribution as a function of 
$\lambda$ for various relaxation rates. Fig.\ \ref{fig3}(b) shows the dependence of $\lambda_0$ on
relaxation time $\tau$ for $\Delta q_a=1$. Note that $\lambda_0$ grows only very slowly with increasing relaxation. While the $\lambda_0$ vs.\ $\tau$ dependence is close to a power law with an exponent $1/4$ - $1/3$, no simple scaling can be expected. The reason is that the scaling suggested by the rate equation (\ref{rateeq}) amended by the relaxation term is incompatible with the scaling implied by the boundary condition $\lambda^{2(\alpha+1)} f'(0)=-\tau_0/\tau S a$ at $q=0$, where $S=1+P^0$. 

\subsection{Additional vibrational modes\label{sec21a}}

Typical molecules have many vibrational modes of different vibrational frequencies. We expect that the scenario of the minimal model is most relevant to molecules whose  lowest frequency mode happens to be weakly coupled.  As this mode becomes highly excited, it would start to mix with other (higher-frequency) modes. In the simplest approximation, we can account for such mode mixing as a channel of direct vibrational relaxation so that the discussion of the previous subsection applies. Indeed, due to this mixing, vibrational energy can be distributed among different modes, which will generally tend to decrease phonon occupation numbers, similar to the vibrational relaxation discussed above. Under these conditions, such a weakly coupled vibrational mode may provide an efficient pathway to ``pump" higher-frequency vibrations.

\subsection{Morse potential and dissociation\label{sec22}}
So far, our considerations were based on the harmonic approximation for the phonon potential. We argue however that wide phonon distributions are not an artefact of this approximation, but also appear for more realistic, anharmonic potentials. As an example, we investigate the effect of weak electron-phonon coupling for the Morse potential 
\be V(x) = D\left[e^{-2\beta(x-x_0)}-2e^{-\beta(x-x_0)}\right],\ee
 where $D>0$ denotes the dissociation energy, $\beta$  the inverse range, and $x_0$ the potential minimum. 

The Morse potential\cite{Morse} accurately describes the vibrations of diatomic molecules and allows us to study current-induced
molecular dissociation.\cite{seideman} This phenomenon has been explored experimentally in STM experiments with molecules on metal surfaces, and sufficiently high currents have been reported to lead to fast dissociation of the absorbed molecules.\cite{stipe} While this scenario with strong coupling between the molecule and the metal surface is distinct from the regime addressed in this paper, we remark that, in principle, the deposition of molecules on passivated surfaces\cite{ho} can be exploited to study the weak-coupling regime as well.

In analogy to the harmonic oscillator model, we assume that the potential energy curves for the neutral and singly-charged molecule have the same shape (i.e.~$D$ and $\beta$ are fixed), but are shifted with respect to each other by $\Delta x=\sqrt{2}\lambda\ell_\text{osc}$. Building on e.g.\ Ref.\ \onlinecite{lemus}, we derive Franck-Condon matrix elements for the 
Morse potential, analogous to Eq.\ \eqref{HO_matrixelement}, between bound states as well as between bound and continuum states. Details of this calculation are referred to Appendix \label{app}.

\begin{figure}
  \centering
		\includegraphics[width=0.7\columnwidth]{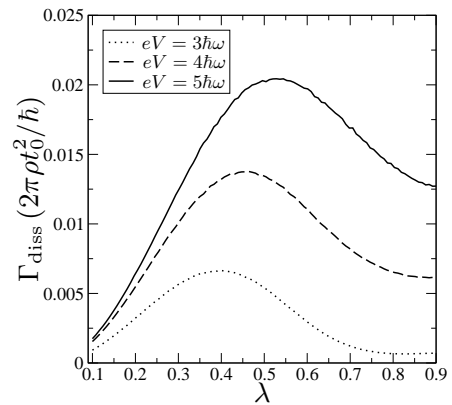}
	\caption{Mean dissociation rate as a function of electron-phonon coupling strength at $k_BT=0.005\hbar\omega$ for several bias voltages as obtained by Monte Carlo simulation with a Morse potential containing 10 bound states. 
 \label{fig4}}	
\end{figure}

Specifically, we study the current-induced dissociation rate of the molecule as function of electron-phonon coupling
and bias voltage by Monte-Carlo simulation. Assuming low temperatures and switching on the voltage at $t=0$, the molecule starts in the phonon ground state and then evolves in time due to the tunneling dynamics. Given that transitions from the continuum back to bound states are negligible, we obtain an average dissociation rate $\Gamma_\text{diss}$ by recording the times $t_{\text{diss},i}$  required for reaching the continuum and averaging over samples. (We note that calculations of mean first-passage times for the highest-lying bound level give compatible dissociation times.) Typical results for dissociation rates for weak electron-phonon couplings (without relaxation) are depicted in Fig.~\ref{fig4}. 

The maximum in the dissociation rate $\Gamma_{\rm diss}$ vs.\ $\lambda$ can be understood as a direct 
consequence of a competition
between the broadening of the phonon distribution and the slowing down of diffusion in
phonon space. As $\lambda$ decreases from values of the order of unity,  
$\Gamma_{\rm diss}$ first increases. This reflects the concurrent increase in the width of the phonon
distribution. Beyond the maximum, $\Gamma_{\rm diss}$ decreases due to the slowing down of diffusion in phonon space. Finally, the 
dissociation rate increases with voltage because of the increased width of the phonon distribution 
(see Fig.\ \ref{fig1}) and the possibility of multiple-phonon excitations within one tunneling event.

\section{Conclusions\label{sec3}}
We have studied the current-induced vibrational nonequilibrium in single-molecule devices
and found that remarkably, the width of the nonequilibrium phonon distribution increases with decreasing 
electron-phonon coupling. We have identified regions in the bias voltage-gate voltage plane in which the width 
of the phonon distribution exhibits power-law divergences with decreasing $\lambda$, with voltage-dependent
non-integer exponents. In some representative cases, we are able to derive analytical phonon distributions 
by a mapping to a Fokker-Planck equation. These striking effects of current-induced nonequilibrium are found
to have important implications in more realistic models which include direct vibrational relaxation and 
anharmonic potential surfaces. A very important conclusion from our work is that approaches which are 
perturbative in the electron-phonon coupling $\lambda$ have to be assessed with extreme care in fully-developed 
nonequilibrium. Finally, we remark that recent experiments\cite{LeRoy} show that the vibrational relaxation 
time can be as large as 10ns, in which case current-induced vibrational nonequilibrium becomes important for 
currents as small as 10pA. 

\begin{acknowledgments}
This work was supported in part by Sfb 658, the Junge Akademie (FvO), Studienstiftung des deutschen Volkes (JK), and
the Israel Science Foundation (AN).
\end{acknowledgments}

\appendix
\section{FC matrix elements for the Morse potential}
The calculation of dissociation rates within the Morse-potential model requires not only the determination of FC matrix elements between bound states [which is straightforward and can be found, e.g., in Ref.\ \onlinecite{koch5}], but also matrix elements between bound and continuum states. Although analytical expressions for the continuum eigenfunctions of the Morse potential are known,\cite{matsumoto} their structure involves confluent hypergeometric functions with complex parameters, rendering a direct numerical evaluation of the FC matrix elements difficult.

Instead, we make use of the complete set of orthonormal functions introduced in Ref.\ \onlinecite{lemus},
\be
\phi_n(x)=\sqrt{\frac{\beta n!}{\Gamma(2\sigma+n)}} L_n^{2\sigma-1}(y) y^\sigma e^{-y/2}.
\ee
Here, we denote $y=(2j+1)e^{-\beta x}$ and $\sigma=j-\lfloor j \rfloor $. $j$ is fixed by the Morse potential parameters to 
\be
2j+1=\sqrt{8\mu D/\beta^2\hbar^2},
\ee
and $\lfloor j \rfloor$ [which is the integer closest to and smaller than $j$] gives the number of bound states. This set of functions has three appealing properties:\cite{lemus} (i) It forms a \emph{discrete} complete orthonormal basis enumerated by $n=0,1,2,\ldots$. (ii) The first $\lfloor j \rfloor+1$ functions form a basis for the bound eigenstates of the Morse potential, all remaining functions span the space of continuum eigenstates. (iii) With respect to this basis, the Hamiltonian takes a particularly simple tridiagonal form.

Denoting the bound and continuum eigenstates of the Morse potential by $\ket{\psi_q}$ and $\ket{\psi_E}$, respectively, we can now calculate the relevant FC matrix element for a transition from a bound state $\ket{\psi_q}$ into a continuum state $\ket{\psi_E}$ by
\begin{align}
M_{q\rightarrow E} &= \boket{\psi_E}{e^{\sqrt{2}\lambda\ell_\text{osc}\frac{d}{dx}}}{\psi_q}\\\nonumber
&= \sum_{m=0}^{\lfloor j \rfloor}\sum_{n=\lfloor j \rfloor +1}^\infty \alpha_{En}^* \beta_{km} \boket{\phi_n}{e^{\sqrt{2}\lambda\ell_\text{osc}\frac{d}{dx}}}{\phi_m}.
\end{align}
Here, the expansion coefficients $\alpha_{En}$ and $\beta_{km}$ for continuum and bound eigenstates with respect to the $\{\ket{\phi_n}\}$ basis are obtained through numerical diagonalization of the Hamiltonian. Finally, the FC matrix for the $\{\ket{\phi_n}\}$ basis are given by
\begin{align}
&\boket{\phi_n}{e^{\sqrt{2}\lambda\ell_\text{osc}\frac{d}{dx}}}{\phi_{n'}}=\frac{(-1)^{n'}(4a)^\sigma
\Gamma(n+n'+2\sigma)}{\sqrt{\Gamma(2\sigma +n)\Gamma(2\sigma+n')n! n'!}}\\\nonumber
&\times\frac{(a-1)^{n+n'}}{(a+1)^{n+n'+2\sigma}} {_2\text{F}}_1\left(-n',-n;-n-n'-2\sigma+1;{\textstyle \frac{(1+a)^2}{(1-a)^2}}\right),
\end{align}
where $a=\exp\left[ -\frac{2\lambda}{\sqrt{\lambda+1}} \right]$ and $_2F_1$ denotes the (Gaussian) hypergeometric function. In practice, one introduces a cutoff for the basis $\{\ket{\phi_n}\}$, leading to a discretization of continuum eigenstates. In order to ensure a sufficiently dense spacing of the spectrum close to the dissociation limit, we take into account $\sim10,000$ basis functions.

\end{document}